\documentclass[11pt]{book}
\usepackage{Wiley-AuthoringTemplate}
\usepackage[authoryear]{natbib}
\begin{document}

\chapter*{Optimal Strategies to Contain and Prevent the Spread of Diseases in Group-structured Populations}

\author[1]{Varun K. Rao}
\author[2]{Siddharth Patwardhan}
\author*[1]{Filippo Radicchi}
 
\setcounter{page}{1}
\address[1]{\orgdiv{Luddy School of
Informatics, Computing, and Engineering}, 
\orgname{Indiana University}, 
\postcode{47405}, \city{Bloomington, IN}, \country{USA}}%

\address[2]{\orgdiv{Center for Science of Science and Innovation, Kellogg School of Management}, \orgname{Northwestern University}, \postcode{60208}, \city{Evanston, IL}, \country{USA}}

\address*{Corresponding Author: Filippo Radicchi; \email{filiradi@iu.edu}}

\maketitle

\section*{Abstract}

The strength and speed of an epidemic process in a multiplex network is affected by how modular the networks in each layer are.  However, it is not obvious how the dynamics of the spreading process change when the group structure and membership varies between layers. In this work, we demonstrate that restructuring the organization of groups within a population can be used as an effective strategy to decrease the severity of an epidemic. Specifically, we show that when group structures are sufficiently correlated outbreaks are longer but milder than for uncorrelated group structures. Further, we find that random immunization decreases the outbreak size and peak infected values to a greater extent than in correlated groups when compared to random immunization in the uncorrelated, normal assignments. Finally, we investigate the effectiveness of targeted immunization strategies on networks with correlated and uncorrelated group structures. We show that immunizing nodes based on their centrality ranking decreases the size of outbreaks compared to random immunization, and that immunizing central nodes in a correlated network helps to decrease the outbreak size more quickly than in uncorrelated networks. 

\section*{Introduction}

The fate of an epidemic process is determined by the structure of the social network where spreading occurs~\citep{Pastor-Satorras2015,vespignani2020modelling,liu2018measurability,soriano2018spreading}. In some real-world networks, for example, those formed by college students and school-going children, the vast majority of contacts of physical proximity are due to group activities/settings~\citep{granell2018epidemic, gomez2018critical,soriano2018spreading}. Realistic models of epidemic spreading are informed by data accounting for the existence of multiple types of group interactions that individuals are exposed to~\citep{broeck2011gleamviz}. This aspect is also accounted for in theoretical models of epidemic spreading~\citep{Granell2013,Gomez2013,granell2014competing,soriano2018spreading,gomez2018critical,granell2018epidemic}.

In all the above studies, the exposure of individuals to multiple types of group interactions is considered an input of the modeling framework. Such a perspective can be, however, changed by treating the exposure of individuals to multiple types of group interactions as a free parameter in a model of epidemic spreading with the goal of understanding the effects that group structure and interrelation among groups have on the fate of an epidemic. In this chapter, we summarize the main finding of our recent publication on the topic~\citep{patwardhan2023epispread}. The focus is on a Susceptible-Infected-Recovered (SIR) epidemic process defined on multiplex networks where the layers represent different settings for social interactions. We specifically consider synthetic multiplex networks with builtin community structure, where we can vary the within-layer strength and the cross-layer correlation of the community structure~\citep{fortunato2010community}. The most important finding is that increasing the strength and/or correlation of the community structure can  (i) dramatically reduce the severity of an epidemic outbreak, both in terms of intensity and duration, as well as (ii) greatly enhance the effectiveness of interventions for disease containment. 
Our work provides an example of 
how microscopic changes in a complex network of social contacts may fundamentally 
impact the macroscopic outcome of an epidemic process. 
This work is therefore highly relevant to making complex adaptive systems (CAS) more resilient 
to disruptions from  epidemics.

\section*{Methods}

\subsection*{Multiplex networks}

Multiplex networks are a generalization of simple networks used to represent relational data consisting of multiple types of interactions between the same entities \citep{bianconi2018multilayer}. 
They are essentially a collection of simple networks sharing the same set of nodes, each imagined on a distinct layer representing a distinct type of interaction.
Multiplexes are useful in modeling various complex systems, including social networks, biological systems, and multimodal transportation systems. In social networks, multiplex networks can model multiple types of interactions, like friendships, professional connections, and shared interests. 
By capturing the multi-faceted nature of real-world systems, multiplex networks provide a meaningful and convenient framework for modeling real complex systems.

\begin{figure}[!htb]
    \centering
    \includegraphics[width=0.4\textwidth]{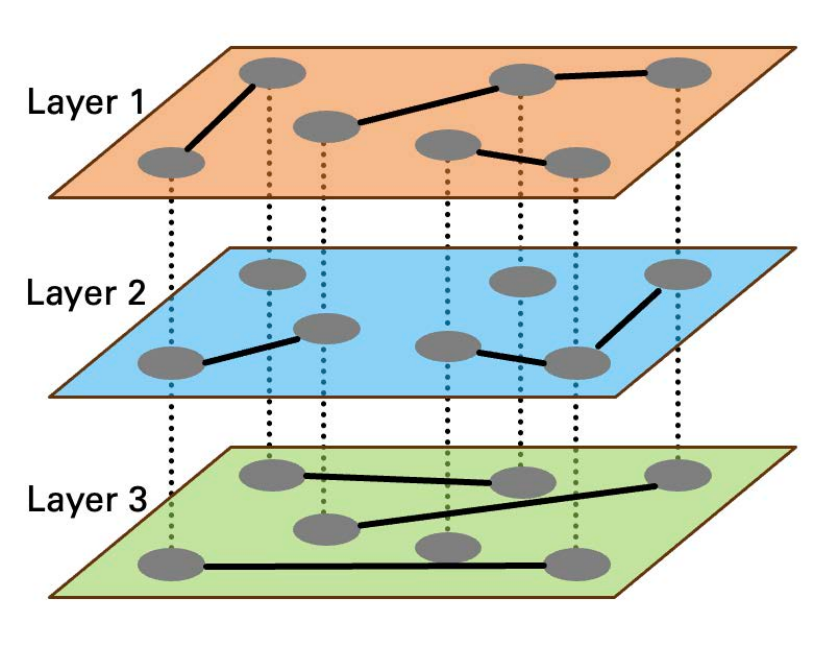}
    \caption{ Visualization of a multiplex network composed of $N=7$ nodes and $L=3$ layers of interactions. Source: \citep{patwardhan2024optimization}.
    }
    \label{fig:multiplex}
\end{figure}

For a multiplex network with $L$ layers, the graph associated with layer $\ell = $1, 2, \ldots, L can be represented as $\mathcal{G}^{(\ell)}=(\mathcal{N},\mathcal{E}^{(\ell)})$. Graphs corresponding to all the layers have the same set of nodes $\mathcal{N}$, with size $N = |\mathcal{N}|$, and have $L$ different layers of interaction given by the edge sets $\mathcal{E}^{(1)},\mathcal{E}^{(2)},...,\mathcal{E}^{(L)}$.
An illustration of a multiplex network with three layers is shown in Figure~\ref{fig:multiplex}. 

A multiplex network can be represented using a set of $L$ symmetric square matrices $\{A^{(1)}, A^{(2)},...,A^{(L)}\}$ of size $N\times N$.
The element $A^{(\ell)}_{i,j}$ of the adjacency matrix represents the relationship between nodes $i$ and $j$ in layer $\ell$.
$A^{(\ell)}_{i,j}=A^{(\ell)}_{j,i}=1$ if there exists an edge connecting $i$ to $j$ in layer $\ell$ and $A^{(\ell)}_{i,j}=A^{(\ell)}_{j,i} = 0$ otherwise.

\subsubsection*{Synthetic multiplex networks with tunable community structure}
In this work, we consider synthetic multiplex networks composed of $L=2$ layers. Each layer has pre-imposed block/community structure and is constructed according to the rules of the stochastic block model, where both degrees and block sizes are homogeneous~\citep{holland1983stochastic}. Specifically,  nodes' degrees in layer $\ell$ obey a Poisson distribution with average  $\langle k^{(\ell)} \rangle$. Also, each individual $i$ is associated to a group $\sigma_i^{(\ell)} = 1, \ldots, Q^{(\ell)}$ in layer $\ell$, with all groups having the same size $q^{(\ell)}$. The mixing parameter $\mu$ quantifies the fraction of connections that each node has towards other nodes outside its own group. We use the same value of the mixing parameter for both layers. We tune the correlation among the community structure of the layers as follows. We start from a configuration where $\sigma_i^{(1)} = \sigma_i^{(2)} \, \forall i=1, \ldots, N$, meaning that the layer-wise community partitions are perfectly correlated. Then, with probability $r$, we swap the label of each node $i$ in layer $\ell = 1$  with that of another randomly selected node $j$. For $r=0$, the two partitions are maximally correlated by construction; for $r = 1$, the two partitions are instead maximally uncorrelated. 
The correlation between the community structure of two layers is measured in terms of normalized mutual information (NMI)~\citep{danon2005comparing}.  Perfectly aligned communities generate large NMI values, whereas uncorrelated block structures have low NMI values.

In some experiments, 
we also consider multiplex networks with heterogeneous degrees and community sizes. There, the degree $k$ of a node is a random variate extracted from the power-law distribution $P(k) \sim k^{-\lambda}$ for $k \in [k_{\min}, k_{\max}]$ and $P(k) = 0 $ otherwise; similarly, the size $q$ of a community is a random variate extracted the power-law distribution $P(q) \sim q^{-\tau}$ for $q \in [q_{\min}, q_{\max}]$ and $P(q) = 0 $ otherwise.

\subsection*{Epidemic Spreading Process}
We consider continuous-time Susceptible-Infected-Recovered (SIR) dynamics on a multiplex network~\citep{Fan2019,bianconi2018multilayer}. 
In the SIR model, nodes can be in three different states: S, I or R. Susceptible nodes can be become infected if in contact with a node in the infected state; infected nodes recover spontaneously; once in the recovered state, nodes do not longer participate in the spreading dynamics~\citep{Pastor-Satorras2015}.

Without loss of generality, we assume that the rate of recovery equals one.
This is a standard setting as the behavior of SIR models is determined by the ratio of the spreading and recovery rates rather than their raw values. We indicate with $\rho_i(t)= S, I$ or $R$ the state of node $i$ at time $t$. Our initial conditions are generally such that all nodes are in the $S$ state, except for a randomly selected node $v$ that is in state $I$, i.e., $\rho_i(t=0) = S$ for $i = 1, \ldots, N$ and $i \neq v$, while $\rho_v(t=0) = I$. In some of our experiments, we also impose that a fraction $f$ of randomly selected nodes is in the recovered state, i.e., $\rho_i(t=0) = R$ with probability $f$ for all $i = 1, \ldots, N$.

Starting from an initial condition, the rules of SIR dynamics are as follows. A generic node $i$ can change its state in the infinitesimal time interval $[t, t+dt]$ if: (i) Node $i$ is in the infected state and recovers with probability $dt$, i.e., $\rho_i(t) = I \to \rho_i(t+dt) = R$ with probability $dt$; 
(ii) Node $i$ is in the susceptible state and gets infected with probability $\beta^{(\ell)} dt$ by its infected neighbor $j$ on layer $\ell$, i.e., $\rho_i(t) = S \to \rho_i(t+dt) = I$ with probability $\beta^{(\ell)} dt$ if $\rho_j(t) = I$ and $A^{(\ell)}_{ij} =1$. Please note that spreading events of type of (ii) can happen for all individual edges in all layers in the graph, thus if node $i$ is in the $S$ state at time $t$, the overall probability to get infected in the time interval $[t, t + dt]$ is $\textrm{Prob.}(\rho_i(t) =S \to \rho_i(t+dt) = I) = 1 - \prod_{\ell=1}^L (1 - \beta^{(\ell)} \, dt)^{g_i^{(\ell)}}$, where  $g_i^{(\ell)} = \sum_j A_{ij}^{(\ell)} \delta (\rho_j(t), I)$ is the number of infected neighbors that node $i$ has in layer $\ell$ at time $t$ with $\delta (x, y) = 1$ if $x=y$ and $\delta (x, y) = 0$ otherwise.
The dynamics proceeds until no nodes are in the infected state. The model is efficiently implemented via the Gillespie's algorithm~\citep{gillespie1977exact}.

\subsubsection*{Metrics of epidemic severity}

We characterize the behavior of the SIR model on a given multiplex network by measuring the values of the following metrics:

\begin{itemize}

\item[(i)] The outbreak size, i.e., the fraction of nodes that are in state $R$ at the end of the epidemic. 

\item[(ii)] The peak of infection, i.e., the maximum value of the fraction of infected nodes that are simultaneously present in the network.

\item[(iii)] The epidemic duration, i.e., the instant of time when the last infected node recovers.

\end{itemize}

Given a multiplex network and the parameters $\beta^{(\ell)}$ of the epidemic model, we simulate the SIR dynamics multiple times. Values of the above metrics displayed in the results of the Chapter correspond to their sample averages. 
\begin{figure}[H]
    \centering
    \includegraphics[width=0.8\textwidth]{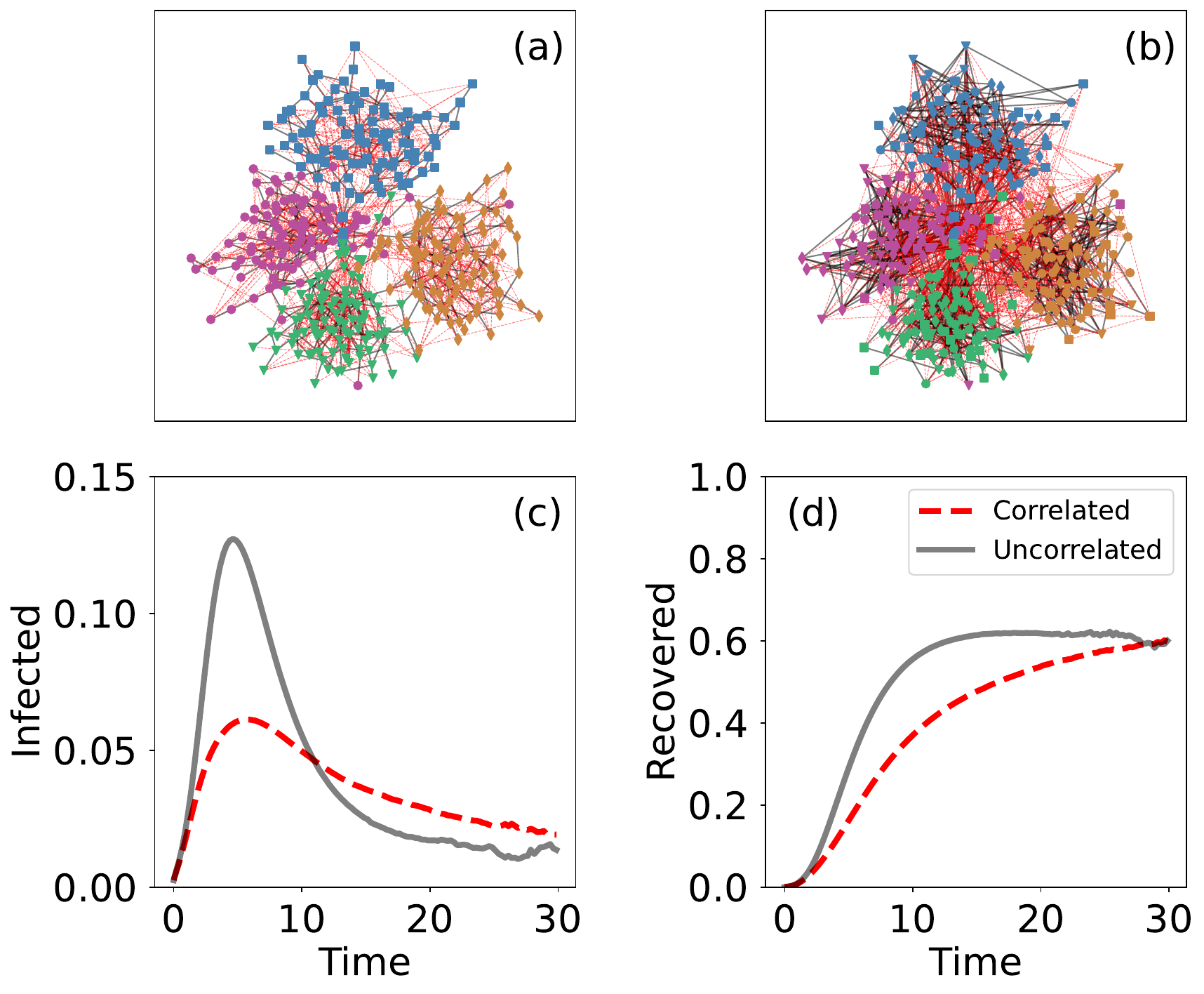}
    \caption{
    (a) To demonstrate the fundamentals of the modeling framework, we consider a multiplex network composed of  
    $N=400$ nodes and two layers of interactions. For both layers ($\ell=1$ and $\ell=2$) nodes are partitioned into four communities  of size $q^{(\ell)} = 100$. Node colors (i.e., orange, purple, blue, green) represent the community structure in layer $\ell=1$; while node shapes (i.e., circle, square, triangle, diamond) indicate the community memberships in layer $\ell=2$. The group structure of these layers are maximally correlated (NMI = $1.0$, with NMI standing for normalized mutual information) within this configuration, meaning that there is a one-to-one map from shapes to colors. Connections between pairs of nodes are created where the average degree in each layer is $\langle k^{(\ell)} \rangle = 3$. A small fraction of these connections connect nodes to other nodes outside their own community, so $\mu = 0.025$ . Using these parameters, we generate edges in layer $\ell=1$ (full gray) and in layer $\ell=2$ (dashed red). (b) Same as in (a), but for uncorrelated community structure (NMI = $0.0$), i.e., colors and shapes are assigned to nodes randomly. (c) We run $V=1,000,000$
    simulations of the SIR model on top of the graphs of panels (a) and (b) by setting the spreading rate 
    $\beta = 0.38$
    We display the fraction of infected nodes in the population as a function of time. We consider bins of size $0.25$ and only report the average values over at least $50$ surviving runs. (d) Same as in (c), but here we are displaying the fraction of recovered nodes in the population as a function of time. 
    }
    \label{fig:0}
\end{figure}
\section*{Results}
\subsection*{Illustrative example}

Fig. ~\ref{fig:0} provides an illustrative example of the modeling framework. All model parameters are fixed, and we generate a small multiplex network in which each layer exhibits a clear block structure, constructed using $\mu = 0.025$. Only the correlation between the community structures of the layers is varied. In Fig.~\ref{fig:0}(a), the two structures are maximally correlated ($NMI =1.0$), whereas in Fig.~\ref{fig:0}(b) they are completely uncorrelated ($NMI =0.0$). Simulations of SIR dynamics on these networks show that, although final outbreak sizes are similar, networks with correlated community structure exhibit milder and longer epidemics than those with uncorrelated groups. This example illustrates that topological correlations among layer-wise community structures influence the epidemic process, while having only a small impact on the final outbreak size.


\subsection*{Group size}

\begin{figure}[!htb]
   \centering
   \includegraphics[width=0.8\textwidth]{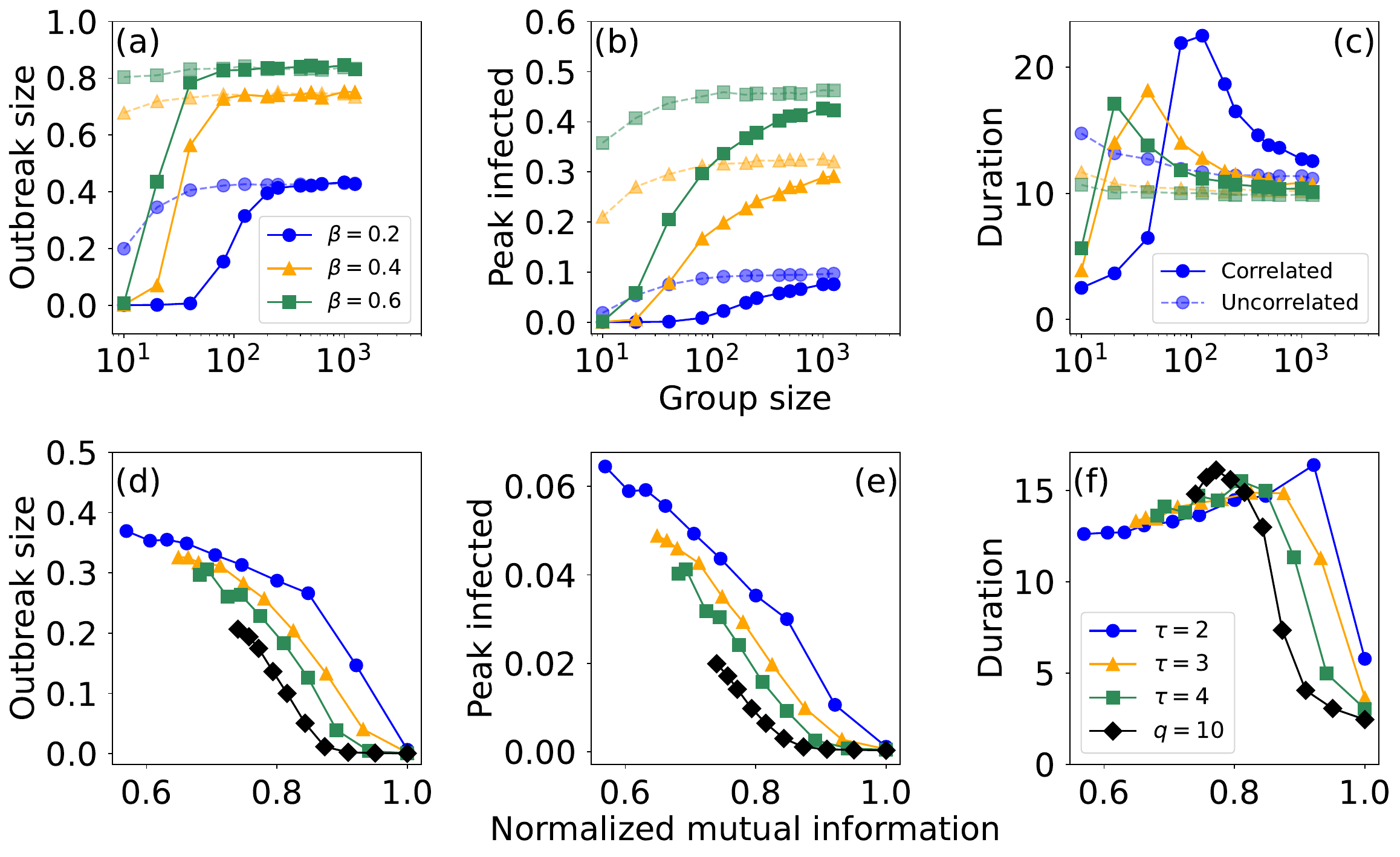} 
   \caption{
   We consider multiplex networks with $N=10,000$ nodes.
   Graphs with $\langle k^{(1)} \rangle = \langle k^{(2)} \rangle = 6$ and $\mu = 0.025$ are considered. We set the community sizes $q^{(1)} = q^{(2)} = q$, and vary $q$ from $10$ to $1250$. 
   The solid and dotted curves show the indicated epidemic properties as a function of the group size $q$ for multiplex networks with correlated and uncorrelated community structures, respectively. 
   Different colors refer to different values of spreading rate $\beta$ in panels a, b, and c.  
   We run $V=5,000$ simulations of the SIR dynamics and display average values in the plot. 
   (a) Outbreak size, (b) fraction
   of the infected population at the peak, and (c) total duration of the outbreak are plotted as functions of the group size.
   Furthermore, we fix the spreading $\beta=0.4$ and choose the community sizes $q$ from the power-law distribution $P(q) \sim q^{-\tau}$ for $q \in [10, 100]$ and $P(q) = 0 $ otherwise. 
   We consider exponent values $\tau = 2$, $3$, and $4$. 
   As a term of comparison, we also display results valid when communities have all identical sizes $q=10$.
   Different colors refer to the different community size distributions in panels d, e, and f.
   (d) Outbreak size, (e) fraction of the infected population at the peak, and (f) total duration of the outbreak are plotted as functions of the normalized mutual information between the layer-wise community structures.
   Source: \citep{patwardhan2023epispread}
   }
   \label{fig:groupsize}
\end{figure}

In Fig.~\ref{fig:groupsize}, we vary the group size in synthetic multiplex networks and show its effect on the metrics of epidemic severity. We look at the outbreak size, the fraction of the population infected during the peak, and the duration in the correlated and uncorrelated multiplex networks as a function of the group size $q$. 
We set the community sizes $q^{(1)} = q^{(2)} = q$ and consider three different values of $q$. We consider graphs with size $N=10,000$, mixing parameter $\mu=0.025$, and average degree $\langle k^{(1)} \rangle = \langle k^{(2)} \rangle = 6$. 

We also explore the effect of the heterogeneous community sizes on the outbreak. We consider multiplex networks with $N = 10,000$ nodes, average degrees $\langle k^{(1)} \rangle = \langle k^{(2)}\rangle = 6$ and $\mu = 0.025$.  We choose the community sizes $q$ from the power-law distribution $P(q) \sim q^{-\tau}$ for $q \in [10, 100]$ and $P(q) = 0 $ otherwise. We consider exponent values $\tau = 2$, $3$, and $4$. 
Community sizes $q_1, q_2, \ldots, q_Q$ are generated from the above power-law distribution and used in both layers to define their block structure.
As a term of comparison, we also display results valid when communities have all identical sizes $q=10$. We tune the correlation among the community structure of the layers by swapping community memberships of nodes as explained in the Methods section. We run $V = 5,000$ simulations of the SIR dynamics and display average values in Figure \ref{fig:groupsize}.


\subsection*{Group correlation and strength}

\begin{figure}[!htb]
    \centering
    \includegraphics[width=0.8\textwidth]{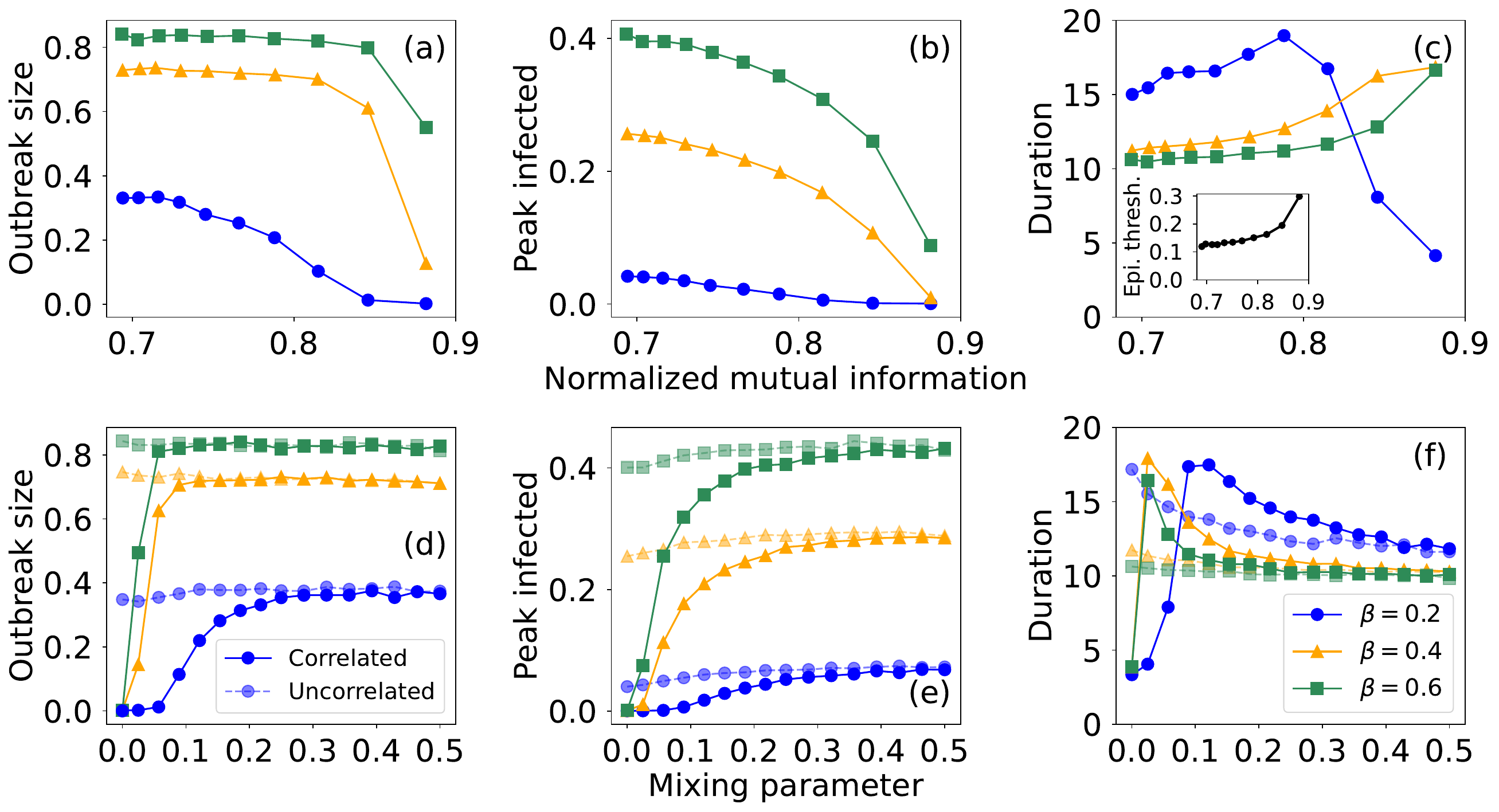}
\caption{
(a) We consider multiplex networks with $N=10,000$ nodes and parameters $\langle k^{(1)} \rangle = 3$, $q^{(1)} = 5$,  $\langle k^{(2)} \rangle = 10$, $q^{(2)} = 25$, and $\mu = 0.025$. We tune the correlation among the community structure of the layers by swapping community memberships of nodes as explained in the Methods section. We simulate SIR dynamics and measure the size of the outbreak. We plot it as a function of the NMI between the layers' partitions. Results are averaged over $V=5,000$ repetitions. Different colors/symbols refer to results valid for different choices of the spreading rate $\beta$, i.e., $\beta =0.2, 0.4$ and $0.6$, all corresponding to supercritical spreading rates [see inset in panel (c)]. The values of the reproduction number are $R_0= 2.6$, $5.2$, and $7.8$, respectively. (b) Same as in (a), but for the peak value of the fraction of infected. (c) Same as in (a), but for the duration of the epidemic.
(d) We consider only the configurations corresponding to maximum (full curves, solid symbols) and minimum (dashed curves, transparent symbols) correlation among layers' partitions and generate networks with variable mixing parameter $\mu$. We plot the size of the outbreak as a function of $\mu$. Results represent averages over $V=5,000$ realizations of the model. (e) Same as in (d), but for the peak of the fraction of infected. (f) Same as in (d), but for the average duration of the epidemic. Source: \citep{patwardhan2023epispread}
    }
    \label{es:fig:3}
\end{figure}

We systematically investigate how correlations between the community structures of multiplex network layers affect SIR spreading. To this end, we apply the algorithm described in the Methods section. Figure~\ref{es:fig:3} reports results for networks with $N=10,000$ and parameters $\langle k^{(1)} \rangle = 3$, $\langle k^{(2)} \rangle = 10$, $q^{(1)} = 5$, $q^{(2)} = 25$, and $\mu = 0.025$. We consider three distinct values of the SIR spreading rate $\beta = 0.2, 0.4$, and $0.6$, all of them corresponding to the endemic regime of the dynamics. We find that increasing group correlation of the layers affects the outbreak severity metrics in unique ways. The final outbreak size remains largely unchanged except when the community structures approach maximal correlation, whereas reductions in epidemic intensity and an increase in epidemic duration are observed over a broader range of correlations. Overall, as the correlation of the groups increases, spreading slows down, and there is a clear reduction in the peak value of the fraction of infected and an apparent increase in the total duration of the epidemic.

These observations hold when network layers have a sufficiently strong modular structure. When the layers' groups are uncorrelated, see Figs.~\ref{es:fig:3}d-f,   there is a negligible effect on the epidemic severity outcomes if the layers' partitions are uncorrelated, . However, if partitions are correlated, we observe increases in all epidemic severity outcomes as the community structure progressively becomes loose.

\subsection*{Immunization}

\begin{figure}[!htb]
    \centering
   \includegraphics[width=0.8\textwidth]{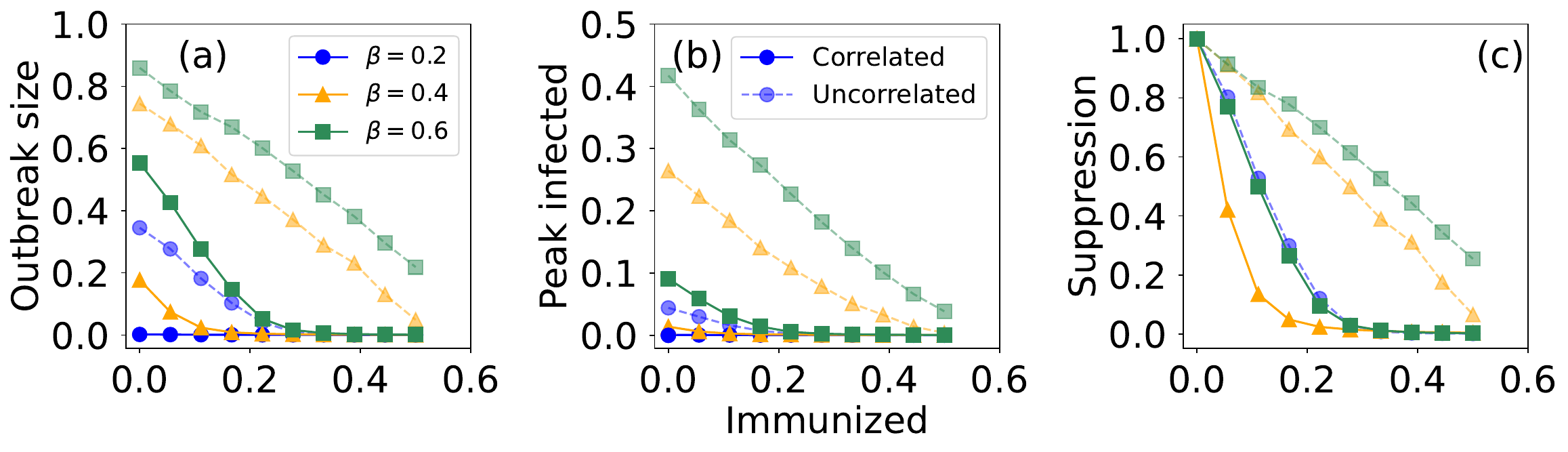}
\caption{
In all panels, we use the same experimental setting as in the bottom row of Fig.~\ref{es:fig:3}, but, modify the initial conditions of the dynamics by immunizing a random fraction of nodes randomly immunize a random fraction of nodes. In the first two panels, we plot the effect this has on two specific outbreak severity outcomes: (a) the outbreak size while in (b) the peak fraction of infected nodes is shown. (c) We shown a novel metric, supression, by rescaling the abscissa values of panel (a) by the outbreak size that is observed when a null fraction of nodes is immunized. All results are obtained by averaging the outcome of $V=5,000$ repetitions of the epidemic process.
Source: \citep{patwardhan2023epispread}
}
    \label{fig:5}
\end{figure}

We find that the effectiveness of immunization in suppressing an epidemic is affected by the two main ingredients of our model in ~\ref{fig:5}, where we  the strength of the community structure of the individual layers ($\mu$), and the correlation among the community partitions in each layer. In ~\ref{fig:5} we explore how varying the level of immunization of the network affects the different outbreak severity metrics and measure how wel

\begin{figure*}[h]
\centering
   \includegraphics[width=0.8\textwidth]{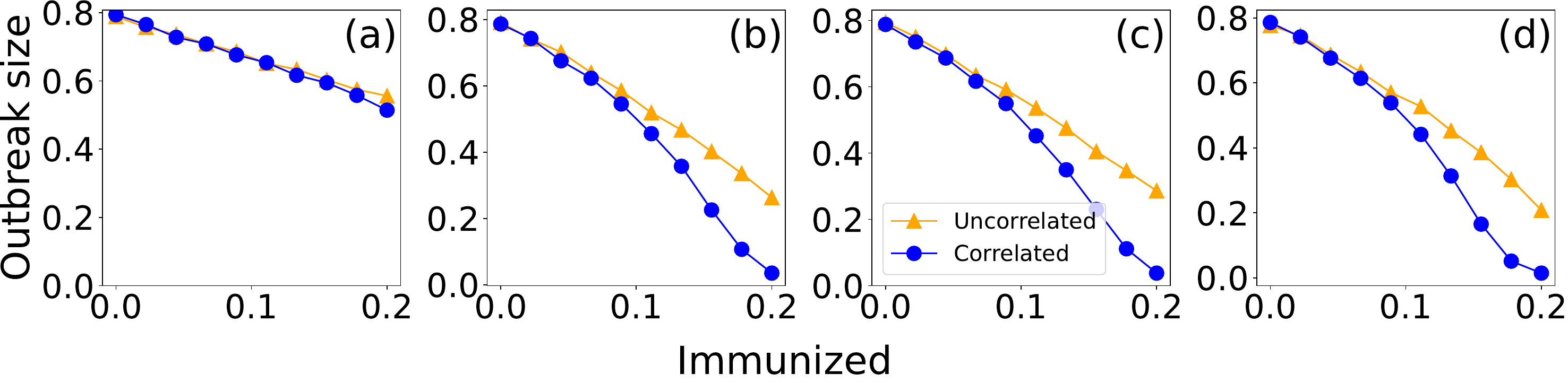}
\caption{
We consider a similar experimental setup, to Fig. ~\ref{fig:5}, with some key differences. In this experiment, we set the spreading rate $\beta = 1$ and utilized a heterogeneous degree distribution. Nodes in the layers are organized in communities of size $q = 100$ and their degrees are chosen from a power-law degree distribution given by $P(k) \sim k^{-2}$ for $k \in [2, \sqrt{q}]$ and $P(k) = 0 $ otherwise. The mixing parameter is $\mu = 0.025$. We consider the cases when the layer-wise community structures are maximally correlated (blue) and when they are uncorrelated (orange). Similar to Fig.~\ref{fig:5},  we immunize a fraction of nodes randomly in (a). Then in (b), (c), and (d) we perform a targeted immunization of the top-ranked nodes based on three different centrality measures: (b) degree, (c) collective influence  (with parameter $\ell=2$)\citep{morone2015influence}, (d) adaptive degree~\citep{chen2009efficient}. We then plot the size of the outbreak as a function of the fraction of immunized nodes. Results are obtained by averaging the outcome of $V=5,000$ repetitions of the epidemic process.}%
   \label{fig:deg_imm}
\end{figure*}
In addition to immunizing nodes based on a random immunization scheme, we test how immunizing nodes based on their centrality measures would affect the size of outbreaks in correlated and uncorrelated networks. In contrast to Fig. ~\ref{fig:5} we use a heterogeneous degree distribution $P(k) \sim k^{-2}$. We utilized a heterogeneous degree distribution to analyze the effects of targeted immunization on the outbreak size to better match real-world systems.  We test the effects of immunizing the nodes with the largest degree, adaptive degree \citep{chen2009efficient}, and collective influence \citep{morone2015influence}. We find that vaccinating nodes based on centrality ranking had a larger effect on decreasing outbreak size than vaccinating nodes randomly. The effect is more pronounced in correlated networks compared to uncorrelated networks.  Immunizing $20\%$ of the nodes by degree, collective influence, or adaptive degree make the outbreak size more than ten times smaller than immunizing the same fraction of nodes randomly in correlated networks.

\section*{Discussion}
In this chapter, we reported on results concerning Susceptible-Infected-Recovered (SIR) spreading in multiplex networks with built-in community structure. 
At the level of individual layers, the modular structure is characterized by the strength and size of the blocks/communities that are present in the layer. 
Correlation among the layer-wise community structures is indicative of the likelihood that two nodes belong to the same community in both layers.

To understand the impact that group structure can have on epidemic spreading, we considered synthetic multiplex networks where the above-listed ingredients can be finely tuned. By means of extensive numerical simulations, we showed that SIR dynamics is characterized by mild and long spreading processes if the communities of the individual layers are neat and the correlation among the layer-wise community structure is sufficiently high. On the contrary, weak and/or uncorrelated community structures generally correspond to short and intense outbreaks. Immunization in multiplex networks with correlated communities is also greatly more effective than in multiplex networks with uncorrelated layer-wise community structures.

Our findings stem from the basic principle that the severity of an epidemic is proportional to the overall level of mixing between individuals of a network. Correlated modular structure in our framework corresponds to redundant interactions among subsets of individuals, which speeds up the propagation of a disease within a subset but also slows down the diffusion across subsets, thus in the overall network.  Although having a radically different implementation, the mechanism of reducing the severity of an epidemic by enhancing group correlation is similar to the one at the basis of the mitigation strategy proposed by Meidan {\it et al.}, consisting of an alternating lockdown policy that involves half of the population at a time~\citep{meidan2020alternating}. 

The root of this work is based on analyzing how a contact network system can be made more resilient to epidemics through changing its network structure. By correlating the community structure between layers, systems experience longer, less severe outbreak while systems that are uncorrelated experience shorter more intense outbreaks. These results highlight how modifying local structural features in a multi-layered network can lead to significant changes in the outcomes of an epidemic process. More broadly, since epidemic spread is an emergent property occurring from the interactions of the network, the results of this chapter show that small structural changes to a CAS can have a large impact on how the system is globally affected by a dynamical process. Our work supports the idea that microscopic changes to interactions within a CAS can have large impact on its resilience to disruption. 

All our results are based on important simplifications concerning the spreading dynamics of realistic diseases as well as the contact networks that are at the basis of the spread of infectious diseases within real-world networks. Nonetheless, we believe they provide some easy-to-implement principles to reduce the severity of real epidemics that managers can take under consideration when planning group activities in schools, colleges, or other large organizations.

\backmatter

\bibliography{references}
\end{document}